\documentclass[]{spie}  

\usepackage{amsmath,amsfonts,amssymb}
\usepackage{CJK}
\usepackage{graphicx}
\usepackage[colorlinks=true, allcolors=blue]{hyperref}
\usepackage{makecell}
\usepackage{threeparttable}
\usepackage{caption}
\usepackage{subcaption}
\usepackage{soul}

\title{Analysis of Laser-Satellite Deconfliction for Astronomical Observatories}

\author[a]{Christoph Baranec*}
\author[b]{Reed Riddle}
\author[c]{Yuhei Takagi}  
\author[d]{Jim Lyke}

\affil[a]{Institute for Astronomy, University of Hawaii at Manoa, Hilo, HI 96720 USA}
\affil[b]{California Institute of Technology, Pasadena, CA 91125 USA}
\affil[c]{Subaru Telescope, National Astronomical Observatory of Japan, Hilo, HI 96720, USA}
\affil[d]{W. M. Keck Observatory, Kamuela, HI 96743, USA}

\authorinfo{* baranec@hawaii.edu; phone 1-808-498-9817}

\begin{document} 
\maketitle

\begin{abstract}
We present an analysis of laser-satellite deconfliction across three astronomical laser observatories for right ascension and declination targets and fixed azimuth and elevation ranges. This study uses new visualization tools to evaluate how different science exposure durations impact the total percentage of available open-time during observing nights. We further assess the operational efficiency of laser keep-out-cone half-angles from the default 2.5 degrees down to the 0.1-degree minimum. We also directly compare the open-time differences of right ascension and declination targets vs. fixed azimuth and elevation ranges. Finally, we analyze historical trends to quantify the growing effect of satellite mega-constellations and the efficacy of specific waiver protocols on ground-based laser astronomy. 
\end{abstract}


\section{INTRODUCTION}
\label{sec:intro}  

U.S. government funded astronomical observatories that propagate lasers into space are required to coordinate activities with the Laser Clearinghouse (LCH) to avoid the inadvertent illumination of critical space assets. Laser systems first need to be registered with LCH and then undergo a qualification process before being approved. During normal operations, a laser owner will send a Program Request Message (PRM) for a specific night that comprises a list of targets to LCH several days before observing. Allowable target types are listed in LCH's Reports Handbook\cite{LCH}. Before the specified night, LCH will return a Program Approval Message (PAM) to the owner which lists the allowable times, with a temporal resolution of 1 s, to propagate their laser at the respective targets. Recent discussions on this topic, with more detailed information on the implementation and logistical issues, can be found in the following. \cite{NOIR2023, 2024IAUGA..32P1201S, Christou2026}.

\begin{figure} [ht]
   \begin{center}
   \includegraphics[width=\textwidth]{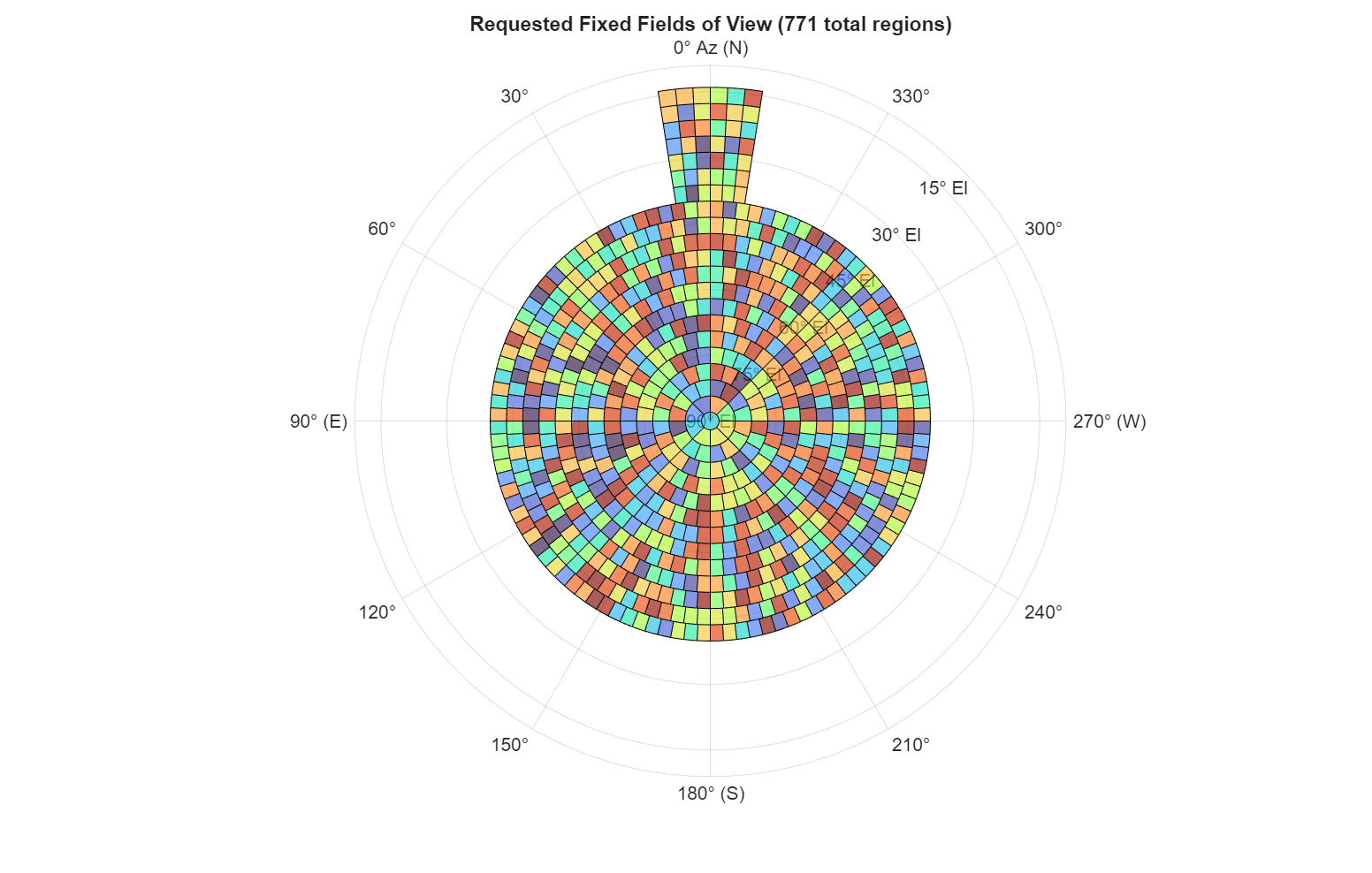}
   \end{center}
   \caption[example] 
   { \label{fig:1} 
Graphical example of Fixed Field of View targets in a Program Request Message submitted to LCH. Each colored region represents a unique, non-overlapping range of Azimuth and Elevation values above the laser observatory.}
   \end{figure}

All astronomical laser observatories used either ``Right Ascension/Declination" or ``Fixed Azimuth/Elevation" target types to describe either scientific or engineering targets respectively, prior to the development of the Robo-AO laser adaptive optics system\cite{Baranec2014}. Robo-AO needs to keep track of tens of thousands of targets every night as part of its intelligent queue, and this ended up being too many targets for LCH to process on a regular basis. During Robo-AO commissioning, LCH imposed a restriction of 6 PRMs per night, each of which may have up to 150 targets, for a total of 900 targets per night. To avoid keeping track of the observability of every target in the Robo-AO queue, we switched to the ``Fixed Field of View" target type which is defined by an Azimuth and Elevation range. This allowed us to effectively tile the sky above the observatory and we only observed scientific targets when they were within a permitted ``Fixed Field of View" (FFoV) target. An example of these targets is shown in (see Fig.~\ref{fig:1}).

\newpage
  
\section{Comparison of PAMs for different laser systems}
\label{sec:observatory_compare}

\subsection{Laser system summary}

The laser observatories and their LCH registered laser parameters evaluated in this study are detailed in Table~\ref{tab:lasers}.

\begin{table}[ht]
\caption{Laser system summary} 
\label{tab:lasers}
\begin{center}
\begin{threeparttable}
\begin{tabular}{|c|c|c|c|c|c|c|} 
\hline
\rule[-1ex]{0pt}{3.5ex}  Laser system & Location & \makecell{$\lambda$ \\ (nm)}  & \makecell{Avg. Power \\ (W)} & \makecell{Peak Power \\ (kW)} & \makecell{Proj. Aperture \\ Diameter (m)}  & \makecell{Focus dist. \\ (km)} \\
\hline
\rule[-1ex]{0pt}{3.5ex}  Keck 1 & Maunakea & 589 & 25 & 0.025 & 0.30 & 90+ \\
\hline
\rule[-1ex]{0pt}{3.5ex}  Keck 2 & Maunakea & 589 & 25 & 0.025 & 0.50 & 90+  \\
\hline
\rule[-1ex]{0pt}{3.5ex}  Robo-AO & \makecell{Palomar\cite{Baranec2014}, \\ Kitt Peak\cite{RAO_KP}} & 355 & 10 & 29\tnote{*} & 0.15 & 10 \\
\hline
\rule[-1ex]{0pt}{3.5ex}  \makecell{Robo-AO\cite{Salama2021}, \\ Robo-AO-2\cite{Baranec2024}} & Maunakea & 355 & 13 & 38\tnote{*} & 0.15 & 10 \\
\hline
\rule[-1ex]{0pt}{3.5ex}  Subaru & Maunakea & 589 & 22 & 0.022 & 0.225 & 90+   \\
\hline 
\end{tabular}
\begin{tablenotes}
    \item[*] 34 ns rectangular pulses at a 10 kHz repetition rate
\end{tablenotes}
\end{threeparttable}
\end{center}
\end{table}

\subsection{Evaluation of observability of targets for different observatories}

We submitted FFoV PRMs for each of Keck 1, Keck 2, Subaru and Robo-AO-2 for the night of UTC 2025 November 21 using the geometry shown in Fig.~\ref{fig:1}. This includes the entire sky above an Elevation $40^\circ$ with a notch to the North. All of the FFoV targets were the same among the different laser systems and span regions that are $3.7^\circ$ in Elevation, and are sized in Azimuth to approximate a square shape. 

   \begin{figure} [ht]
   \begin{center}
   \includegraphics[width=300pt]{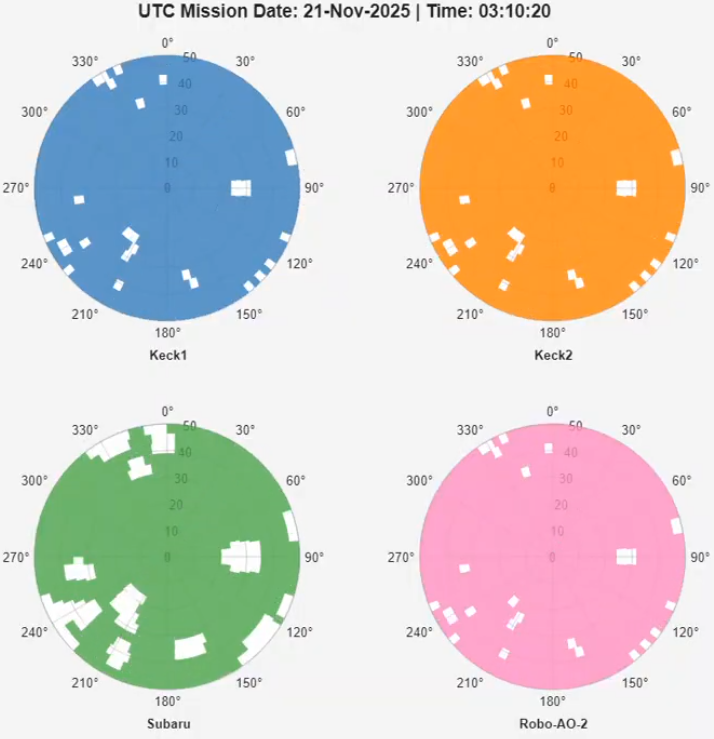}
   \end{center}
   \caption[example] 
   { \label{fig:2} 
Snapshot of the status of the Fixed Field of View targets in the Program Approval Messages for Keck 1, Keck 2, Robo-AO-2 and Subaru. Regions are colored where it is safe to propagate the laser and are shown in white where lasing is prohibited.}
   \end{figure} 

Using the PAMs returned from LCH, we can plot where it is safe (open) and not safe (closed) to propagate the respective lasers in their FFoV regions at every second (see Fig.~\ref{fig:2}). This can then be turned into a movie, which allows for the visualization of the status of the FFoV regions as a function of time. Although one can calculate the fraction of time that any given FFoV region is open, it is of limited value. For example, a region that alternates between open and closed every other second will have the same fractional open time as a region that is closed for the first half of a night, and open for the second half, yet the former would prohibit any real-world observing. Therefore, we calculate the fractional time the regions are open assuming representative observation durations can be executed in an open window: 300 s (typical of imaging) and 1800 s (more typical of spectroscopy). We acknowledge that astronomical targets will not necessarily stay within a single FFoV region during these longer observations due to the sidereal rate, but this still provides us with a more useful comparison. We then calculated the median of all of the fractional open times for FFoV regions to get an estimate of the percent open time for a given observation duration for each laser system for each night. These percentages are presented in Table~\ref{tab:fractions})

\begin{table}[ht]
\caption{Median fractional open time percentages of all regions on UTC 2025 November 21} 
\label{tab:fractions}
\begin{center}
\begin{tabular}{|c|c|c|c|c|} 
\hline
\rule[-1ex]{0pt}{3.5ex}  \makecell{Observation \\ Duration} & Keck 1 & Keck 2  & Robo-AO-2 & Subaru \\
\hline
\rule[-1ex]{0pt}{3.5ex}  1 s & 96.3\% & 96.3\% & 96.4\% & 88.4\% \\
\hline
\rule[-1ex]{0pt}{3.5ex}  300 s & 88.5\% & 88.5\% & 89.8\% & 74.8\% \\
\hline
\rule[-1ex]{0pt}{3.5ex}  1800 s & 30.0\% & 29.9\% & 34.2\% & 11.7\% \\
\hline 
\end{tabular}
\end{center}
\end{table}

From the results in Table~\ref{tab:fractions}, we can see some initial trends. As observation durations increase, the open fraction decreases -- especially so for the 1800 s observations. The fractional open percentage is nearly identical between Keck 1 and Keck 2, which is not surprising given their similar configurations and locations. Despite the very different laser configuration, Robo-AO-2 has a comparable open fraction compared to the Keck telescopes. Subaru had a significantly lower open fraction percentage compared to the other laser systems. A hint of this can be seen in Figure~\ref{fig:2} where Subaru has the same areas of the sky that are closed as the other systems, but it appears the closed regions have been convolved with a kernel of several degrees. We tracked this down to an excessive Keep-Out-Cone on the Subaru laser system; this will be discussed further in the next section.

\section{Effect of laser keep-out-cone half-angle on laser deconfliction}

One element of the laser-satellite deconfliction process that is not integrated into the laser registration is that of determining the keep-out-cone half-angle for a laser system. While the exact method used by LCH for deconfliction is not published, it does take into account error circles and safety margin on both space assets and the laser systems. Presumably a target closure will occur where these error circles overlap. Registered laser systems by default are assigned a 2.5 degree keep-out-cone half-angle by the LCH. Additional evaluation and qualification of the laser system by LCH can be used to determine that a laser system can use a smaller keep-out-cone half-angle, with a minimum of 0.1 degree -- the smallest angle that can be accepted by the LCH deconfliction process.

\begin{figure} [ht]
   \begin{center}
   \includegraphics[width=300pt]{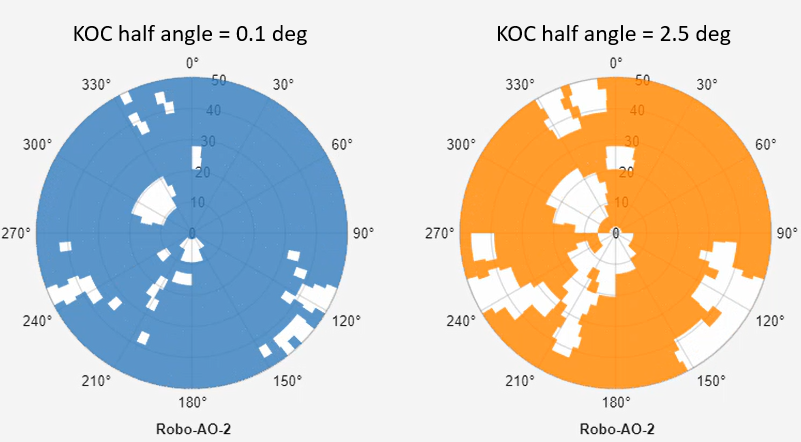}
   \end{center}
   \caption[example] 
   { \label{fig:3} 
Snapshot of the status of the Fixed Field of View targets in the Program Approval Messages for Robo-AO-2 with keep-out-cone half-angles of 0.1 and 2.5 degrees. Regions are colored where it is safe to propagate the laser and are shown in white where lasing is prohibited.}
   \end{figure}

\begin{figure}[htbp]
    \centering
    \begin{subfigure}[b]{0.45\textwidth}
        \centering
        \includegraphics[width=\linewidth]{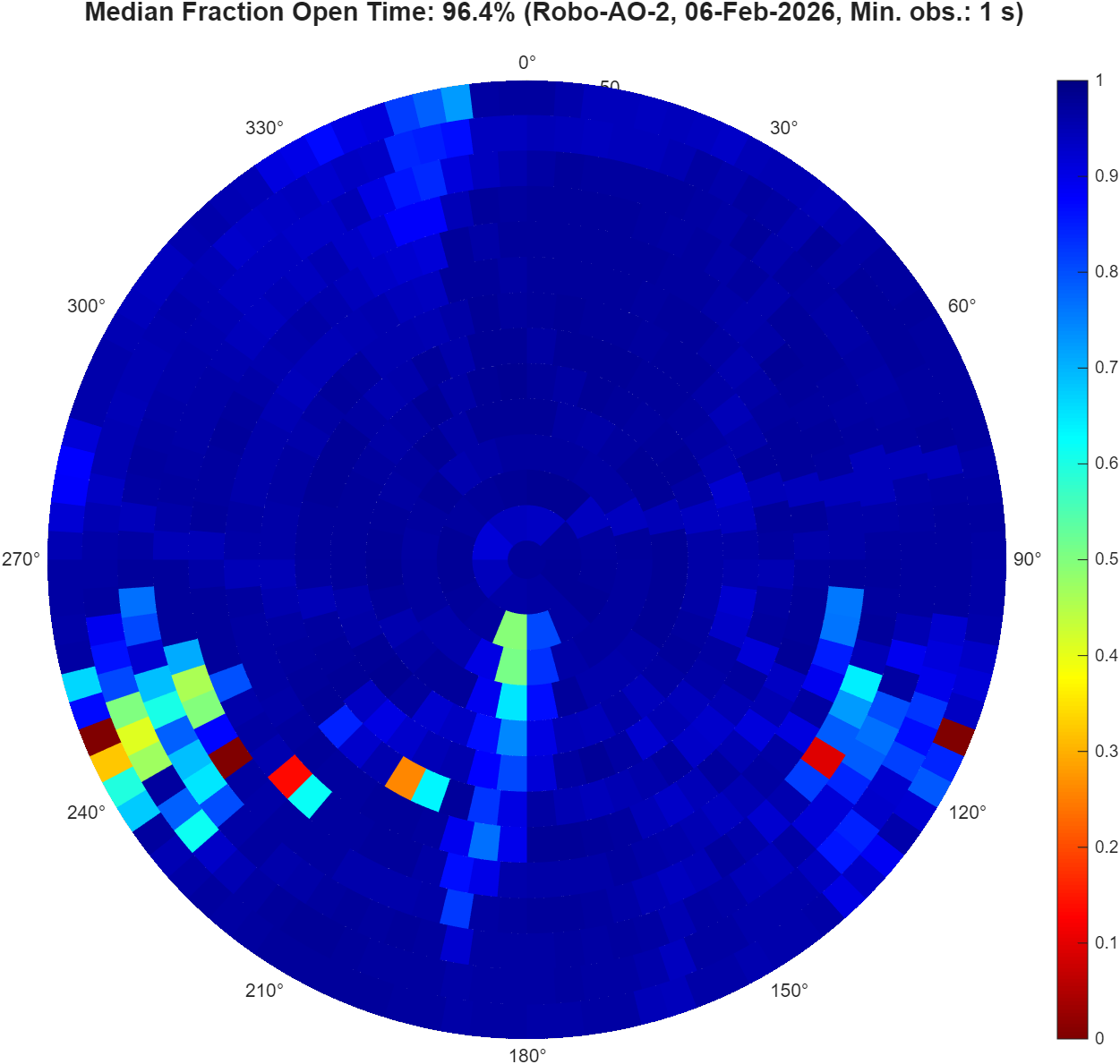}
        \caption{Keep-out-cone half-angle = $0.1^\circ$}
        \label{fig:4-1}
    \end{subfigure}
    \hfill
    \begin{subfigure}[b]{0.45\textwidth}
        \centering
        \includegraphics[width=\linewidth]{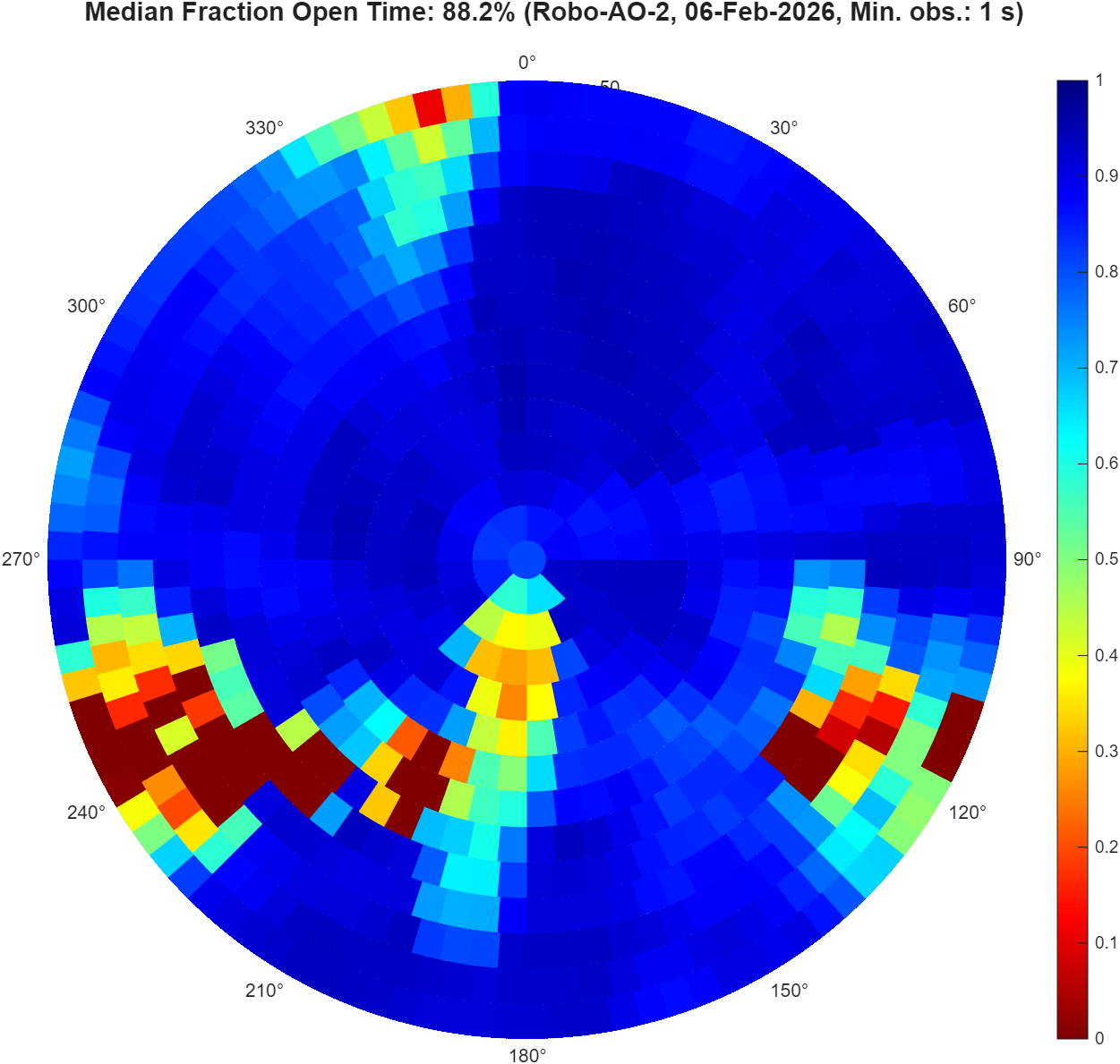}
        \caption{Keep-out-cone half-angle = $2.5^\circ$}
        \label{fig:4-2}
    \end{subfigure}
    \caption{Fractional open time for each FFoV target region using an observation time of 1 s.}
    \label{fig:4}
\end{figure}

\begin{figure}[htbp]
    \centering
    \begin{subfigure}[b]{0.45\textwidth}
        \centering
        \includegraphics[width=\linewidth]{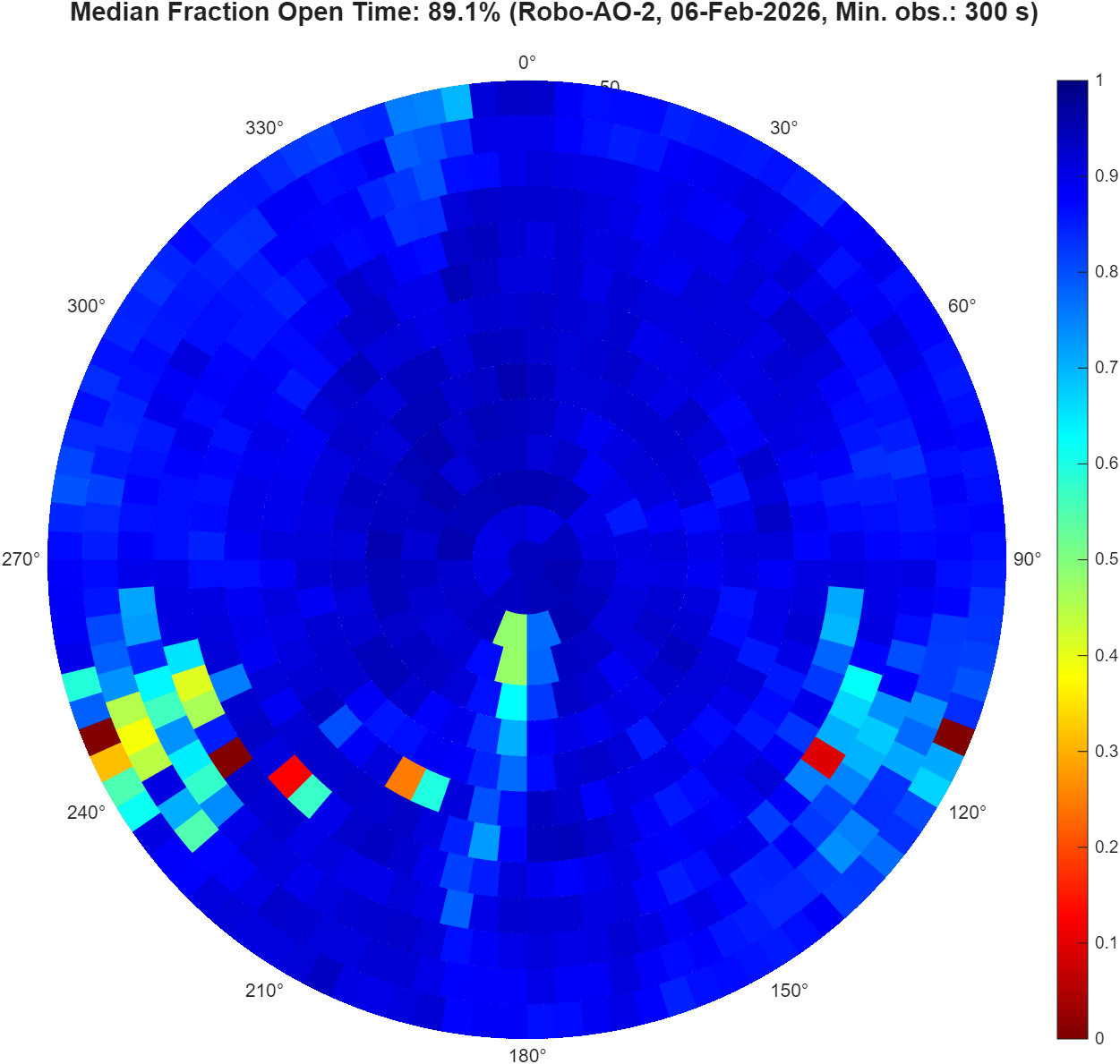}
        \caption{Keep-out-cone half-angle = $0.1^\circ$}
        \label{fig:5-1}
    \end{subfigure}
    \hfill
    \begin{subfigure}[b]{0.45\textwidth}
        \centering
        \includegraphics[width=\linewidth]{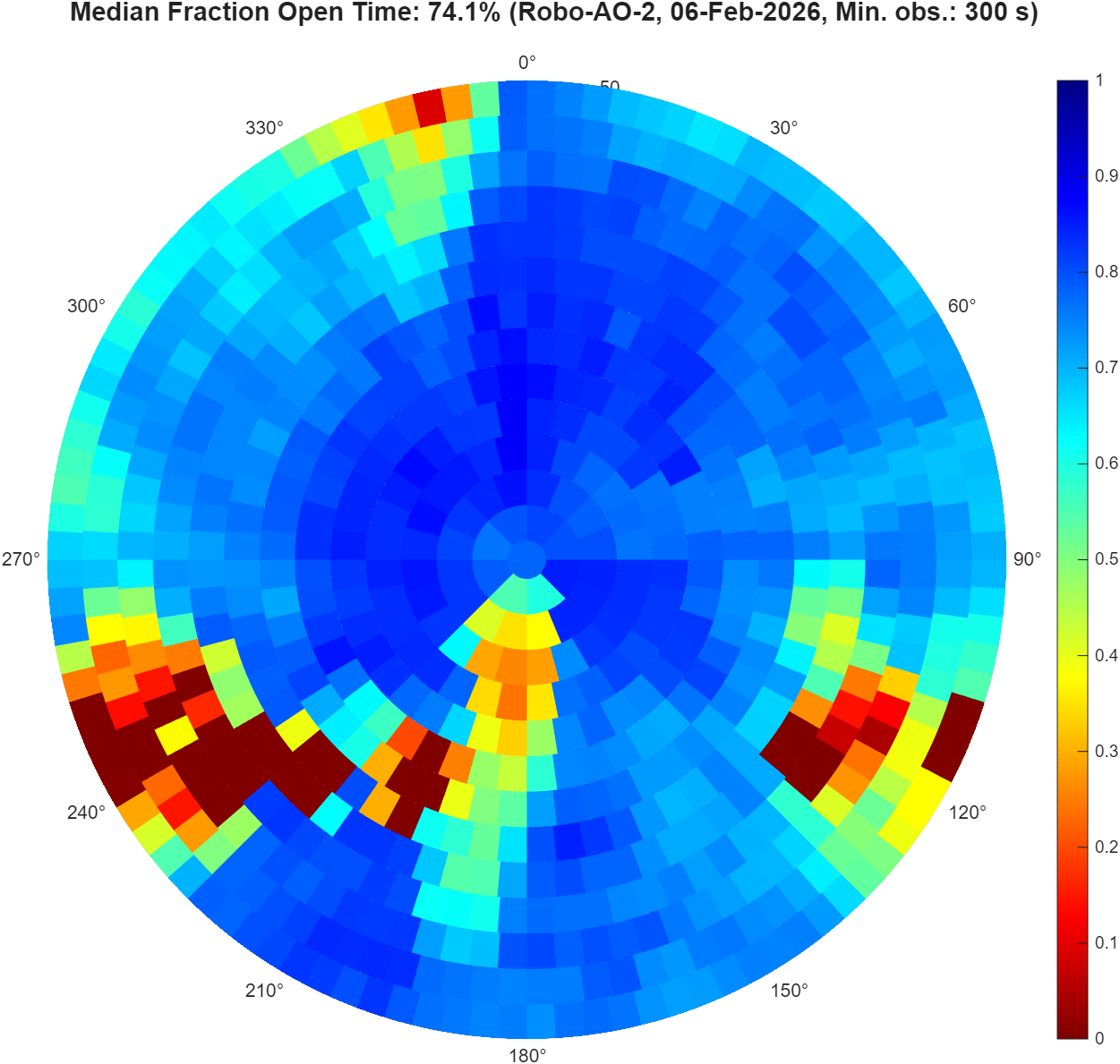}
        \caption{Keep-out-cone half-angle = $2.5^\circ$}
        \label{fig:5-2}
    \end{subfigure}
    \caption{Fractional open time for each FFoV target region using an observation time of 300 s.}
    \label{fig:5}
\end{figure}

\begin{figure}[htbp]
    \centering
    \begin{subfigure}[b]{0.45\textwidth}
        \centering
        \includegraphics[width=\linewidth]{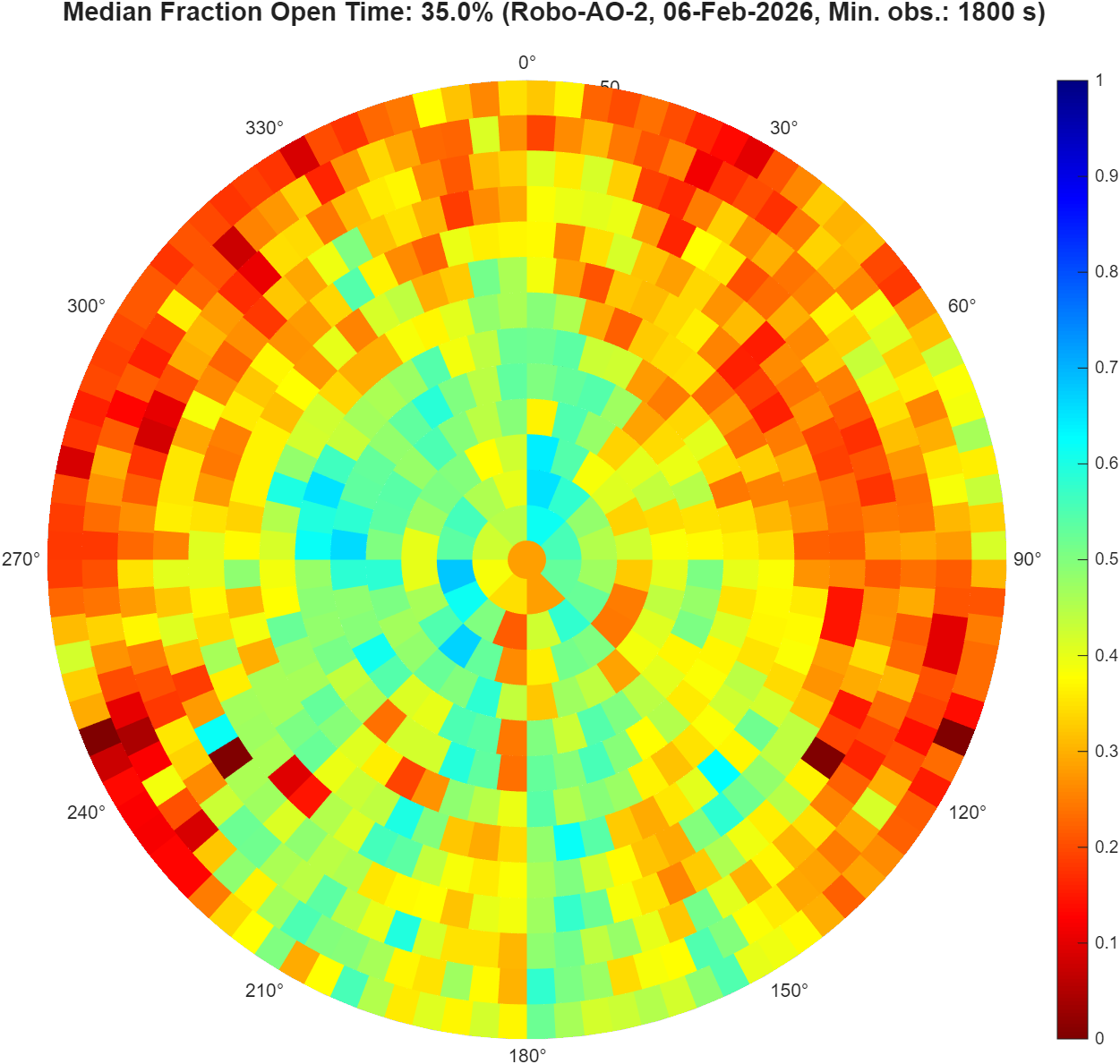}
        \caption{Keep-out-cone half-angle = $0.1^\circ$}
        \label{fig:6-1}
    \end{subfigure}
    \hfill
    \begin{subfigure}[b]{0.45\textwidth}
        \centering
        \includegraphics[width=\linewidth]{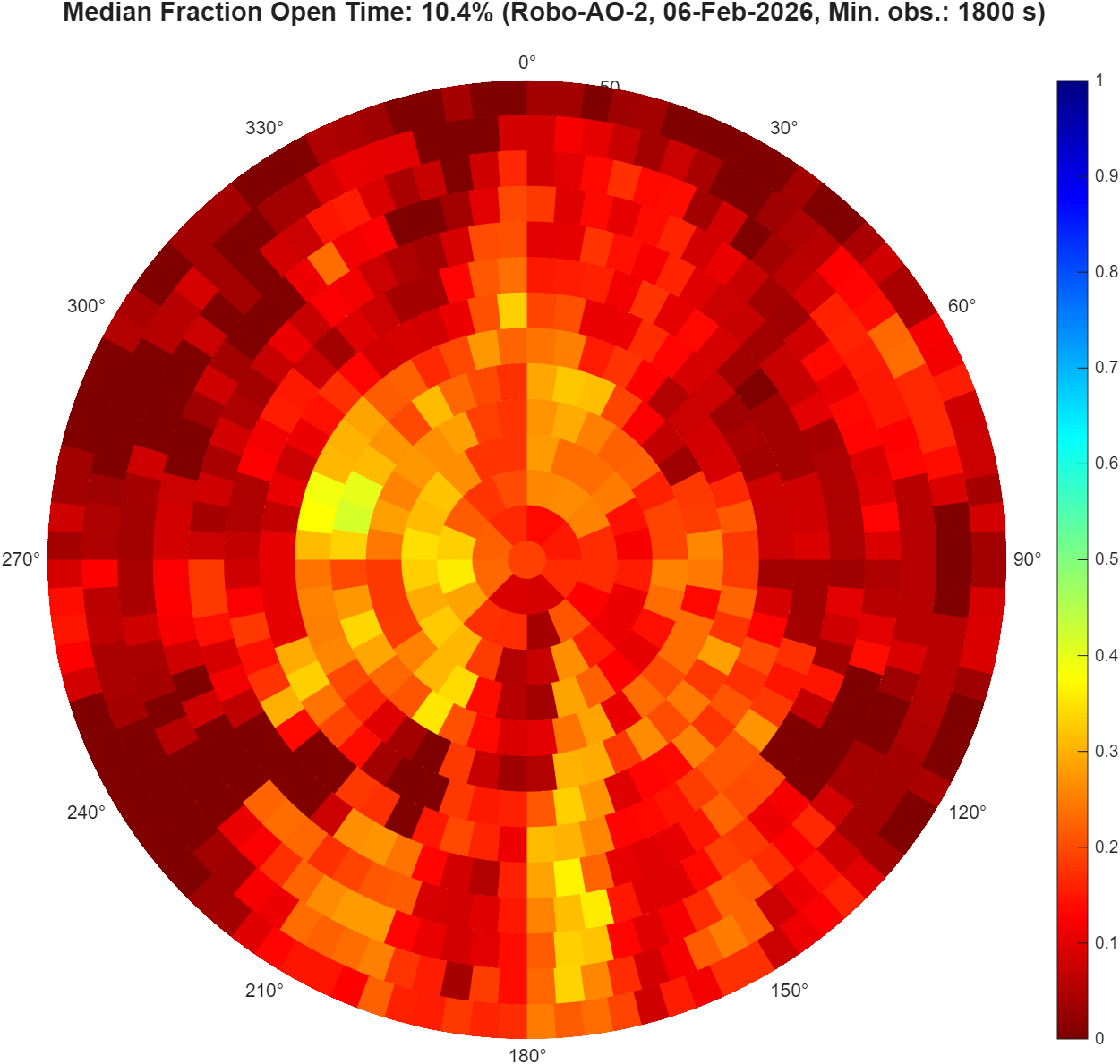}
        \caption{Keep-out-cone half-angle = $2.5^\circ$}
        \label{fig:6-2}
    \end{subfigure}
    \caption{Fractional open time for each FFoV target region using an observation time of 1800 s.}
    \label{fig:6}
\end{figure}

We had the opportunity to request PAMs for Robo-AO-2 on UTC 2026 February 6 that used keep-out-cone half-angles of 0.1 and 2.5 degrees to directly compare the difference (see Fig.~\ref{fig:3}). Over the course of the night, we can determine the fractional open time for each FFoV region using different observation durations (see Figs.~\ref{fig:4}-\ref{fig:6}) and determine the median fractional open time over all of the FFoV regions (see Tab.~\ref{fig:3}).

We see the effect of the larger keep-out-cone half-angles in both the instant FFoV open windows and in the median fractional open time. The median fractional open time values in keep-out-cone half-angles in Table~\ref{tab:koc_fractions} are very similar to the set calculated in Table~\ref{tab:fractions}; with an angle of 0.1 degree correlating with Keck 1, 2 and Robo-AO-2 and with the 2.5 degree angle correlating with Subaru. After this revelation, we suspected that the keep-out-cone half-angle for Subaru had never been changed from the default 2.5 degrees, and then confirmed this was the case. We've since started the process to update the keep-out-cone half-angle used with the Subaru laser system.

What also stands out in Figures~\ref{fig:4}-\ref{fig:6} is there are FFoV regions of the sky, located around a declination of  $0^\circ$, where the laser is never permitted to be used. The presumed space assets in these regions either don't move or move very minimally on sky which is indicative of an object closer to geosynchronous orbit ($\sim$36 Mm) as opposed to low-earth orbit ($<2$ Mm). For scientific programs with targets around this declination, it is critical to be using the smallest keep-out-cone half-angle as possible. Additionally, for longer duration observations, the fractional open time is severely restricted over the entire sky, with only a slight increase in this fraction for FFoV regions that are overhead. 

\begin{table}[ht]
\caption{Median fractional open time percentages of all regions on UTC 2026 February 6} 
\label{tab:koc_fractions}
\begin{center}
\begin{tabular}{|c|c|c|} 
\hline
\rule[-1ex]{0pt}{3.5ex}  Observation Duration & \multicolumn{2}{|c|}{Keep-out-cone half-angle}  \\
\hline
\rule[-1ex]{0pt}{3.5ex}   & $0.1^\circ$ & $2.5^\circ$  \\
\hline
\rule[-1ex]{0pt}{3.5ex}  1 s & 96.4\% & 88.2\%  \\
\hline
\rule[-1ex]{0pt}{3.5ex}  300 s & 89.1\% & 74.1\%  \\
\hline
\rule[-1ex]{0pt}{3.5ex}  1800 s & 35.0\% & 10.4\%  \\
\hline 
\end{tabular}
\end{center}
\end{table}

\section{Comparison of RA/DEC and FFoV regions}

Here we directly compare the difference in open fraction for astronomical targets when using either Right Ascension/Declination targets or FFoV target regions for laser-satellite deconfliction (and using a $0.1^\circ$ keep-out-cone half-angle). We started with a list of simulated astronomical targets in rings of declination separated by $10^\circ$ in declination. The number of targets per declination ring is weighted by the cosine of the declination, and they are evenly distributed in right ascension. We submitted an RA/Dec PRM with these targets to be contemporaneous with the FFoV PAMs for Robo-AO-2 on UTC 2026 February 6.

\begin{figure} [ht]
   \begin{center}
   \includegraphics[width=300pt]{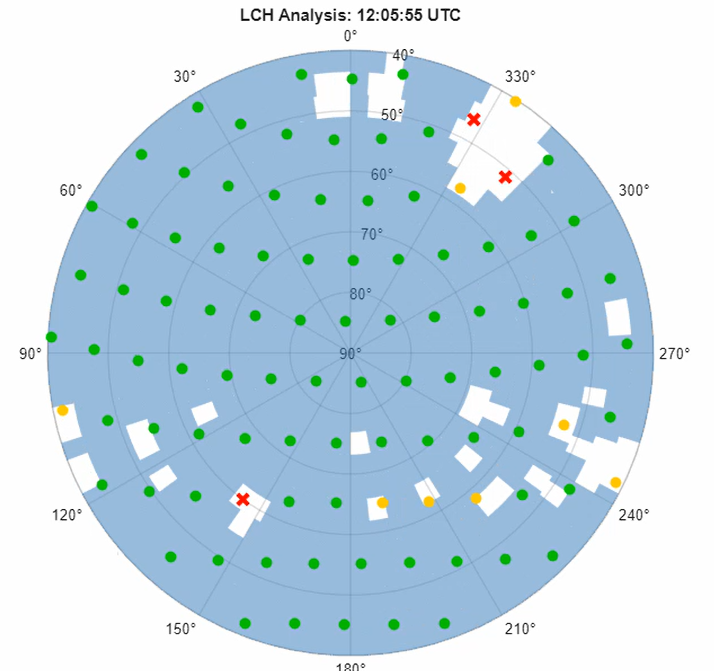}
   \end{center}
   \caption[example] 
   { \label{fig:radeccompare} 
Snapshot of the comparison of PAMs using Right Ascension/Declination targets and FFoV target regions. Regions are colored blue where it is safe to propagate the laser and are shown in white where lasing is prohibited. Simulated astronomical targets are shown as points: green circles where both the RA/Dec and FFoV are open; yellow circles where the RA/Dec PAM deems the target open, but the FFoV views it as closed; and red 'x' makes where the target is closed using both the RA/Dec and FFoV PAMs.}
   \end{figure}

For the analysis, we directly measure the open time fraction for each RA/Dec target given a specific observing duration. We then also generated a simulated RA/Dec PAM file using the FFoV PAM as the source material; for every target at every second we checked which FFoV region the target was in and noted whether it was open or closed. We then repeated the measurement on the open time fraction for each RA/Dec target using this simulated PAM file. A graphical representation of this can be seen in Figure~\ref{fig:radeccompare} showing the FFoV regions and the RA/Dec targets as points with their open status indicated. 

We again calculate the median fractional open time for the astronomical targets using either the RA/Dec or FFoV PAMs with the addition of discriminating by declination. This appears in Table~\ref{tab:radec}. In all cases, the fractional open time for the astronomical targets is higher using the RA/Dec PAM vs the FFoV PAM which can be explained by the fact that the latter includes larger areas on sky and would thus be more prone to closures. While there is not a strong dependence with declination, the effects of the nearly static closures seen in Figure ~\ref{fig:4} show up in the reduced open fraction of the FFoV PAM at $-5^\circ$; presumably this does not show up in the RA/Dec PAM as geosynchronous satellites are nominally at a declination of $-3.4^\circ$ with respect to Maunakea.

\begin{table}[ht]
\caption{Median fractional open time percentages by declination and target type} 
\label{tab:radec}
\begin{center}
\begin{tabular}{|c|c|c|c|c|c|c|c|c|} 
\hline
\rule[-1ex]{0pt}{3.5ex}  Observation Duration $\rightarrow$ & \multicolumn{2}{|c|}{1 s} & \multicolumn{2}{|c|}{60 s} & \multicolumn{2}{|c|}{300 s} & \multicolumn{2}{|c|}{1800 s} \\
\hline
\rule[-1ex]{0pt}{3.5ex}  Declination $\downarrow$ & RA/Dec & FFoV & RA/Dec & FFoV & RA/Dec & FFoV & RA/Dec & FFoV   \\
\hline
\rule[-1ex]{0pt}{3.5ex}  All & 99.0\% & 96.5\% & 98.9\% & 95.9\% & 97.3\% & 88.4\% & 74.0\% & 30.6\% \\
\hline
\rule[-1ex]{0pt}{3.5ex}  $+65^\circ$ & 98.7\% & 89.6\% & 98.6\% & 88.9\% & 97.2\% & 84.3\% & 65.5\% & 15.5\% \\
\hline
\rule[-1ex]{0pt}{3.5ex}  $+55^\circ$ & 98.7\% & 95.9\% & 98.6\% & 95.2\% & 97.2\% & 89.1\% & 69.5\% & 21.9\% \\
\hline
\rule[-1ex]{0pt}{3.5ex}  $+45^\circ$ & 98.9\% & 97.3\% & 98.9\% & 96.9\% & 97.4\% & 92.1\% & 73.1\% & 30.2\% \\
\hline
\rule[-1ex]{0pt}{3.5ex}  $+35^\circ$ & 98.9\% & 97.7\% & 98.8\% & 97.1\% & 97.9\% & 91.6\% & 75.8\% & 32.1\% \\
\hline
\rule[-1ex]{0pt}{3.5ex}  $+25^\circ$ & 98.9\% & 97.9\% & 98.9\% & 97.5\% & 97.2\% & 90.4\% & 77.0\% & 38.5\% \\
\hline
\rule[-1ex]{0pt}{3.5ex}  $+15^\circ$ & 99.1\% & 97.7\% & 99.1\% & 97.3\% & 98.3\% & 90.6\% & 70.6\% & 31.7\% \\
\hline
\rule[-1ex]{0pt}{3.5ex}  $+5^\circ$ & 99.0\% & 91.3\% & 99.0\% & 90.7\% & 96.8\% & 81.7\% & 76.4\% & 30.1\% \\
\hline
\rule[-1ex]{0pt}{3.5ex}  $-5^\circ$ & 98.8\% & 83.0\% & 98.8\% & 82.2\% & 97.5\% & 73.6\% & 76.6\% & 23.8\% \\
\hline
\rule[-1ex]{0pt}{3.5ex}  $-15^\circ$ & 98.8\% & 95.5\% & 98.8\% & 94.4\% & 97.2\% & 87.1\% & 74.4\% & 33.0\% \\
\hline
\rule[-1ex]{0pt}{3.5ex}  $-25^\circ$ & 99.1\% & 97.5\% & 99.1\% & 96.9\% & 97.3\% & 90.4\% & 76.1\% & 35.6\% \\
\hline

\end{tabular}
\end{center}
\end{table}

\section{Analysis of historical FFoV regions}

\begin{figure} [ht]
   \begin{center}
   \includegraphics[width=\textwidth]{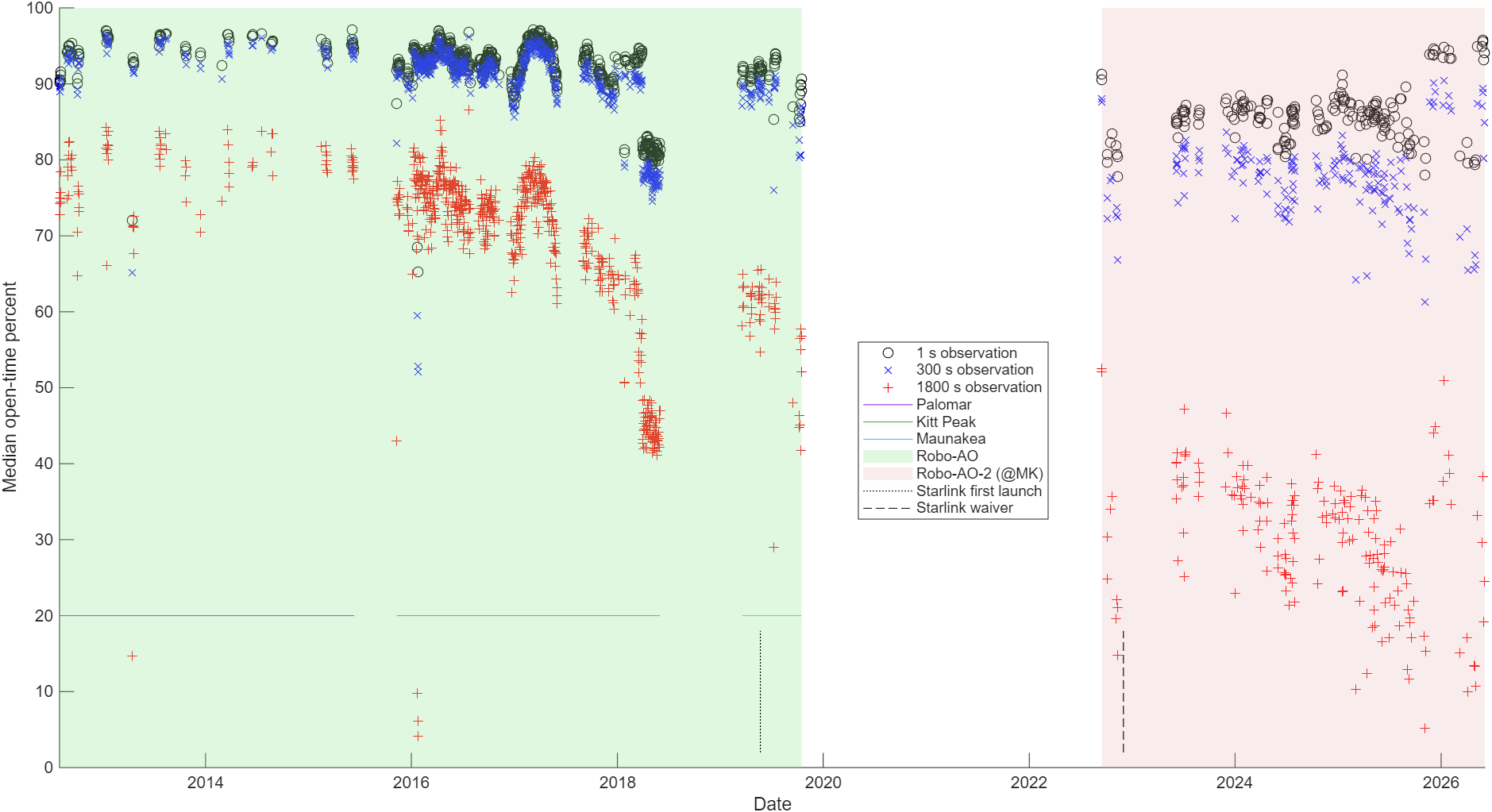}
   \end{center}
   \caption[example] 
   { \label{fig:historical} 
Median fractional open time of FFoV PAMs from the Robo-AO and Robo-AO-2 systems for different observing durations.}
   \end{figure}

We now analyze historical trends using FFoV PAMs from the Robo-AO and Robo-AO-2 system which date back to 2012 August 1. For every night, we calculate the percentage of open time for each FFoV region given an observation and then calculate the median of all of these values. We then plot these points in Figure~\ref{fig:historical}. The location of the Robo-AO system is indicated on the plot: at Palomar from 2012 August 1 to 2019 October 16; at Kitt Peak from 2015 November 10 to 2018 June 01; and at Maunakea from 2019 March 20 to 2019 October 16. Robo-AO-2 is located at Maunakea, starting 2022 September 15. Both Robo-AO and Robo-AO-2 used the default $2.5^\circ$ keep-out-cone half-angle until it was updated to $0.1^\circ$ starting 2025 November 21. There were six nights from March-May 2026 where LCH inadvertently returned Robo-AO-2 to a $2.5^\circ$ half-angle, which was subsequently rectified. Also indicated is the date of the first launch of Starlink satellites on 2019 May 23, as well as the date that Starlink satellites were waived from the protect list for astronomical lasers at the beginning of 2022 December\cite{NOIR2023}.

The open windows appear relatively stable until 2016 when they start to trend lower. The effect is more pronounced for longer observation durations. Unfortunately there is minimal data during the Starlink period where their satellites were automatically included in every protect list so it is difficult to say much conclusively during this time. There are hints that the waiver did increase the median open fraction as there is a slight change between the values calcuated before the waiver and those in mid-2023. The more recent Robo-AO-2 PAMs still show that the median fractional open time is continuing to decrease for all observation durations. There are two groups of data points which do not follow this trend, and they correspond to LCH using the smaller keep-out-cone half-angle; these are qualitatively similar to the values calculated in 2018-2019.   

\section{DISCUSSION}

From the presented analysis, we find that the effect of expanding satellite constellations and specific operational parameters, specifically the chosen target type and keep-out-cone half-angle, has a profound effect on the efficiency of astronomical laser observatories. As the low-Earth orbit grows increasingly congested, optimizing coordination with the Laser Clearinghouse is critical to maintaining scientific productivity.

Based on our findings, we can draw several key conclusions regarding laser-satellite deconfliction strategies and system configurations:

\begin{itemize}
    \item \textbf{Keep-Out-Cone Optimization:} The keep-out-cone half-angle is a dominant factor in determining laser observability. Systems operating at the default $2.5^{\circ}$ half-angle experience severe reductions in open time, as evidenced by the Subaru laser system and our Robo-AO-2 comparative analysis. Reducing this angle to the LCH minimum of $0.1^{\circ}$ is essential to preserve viable observation windows, particularly for longer exposures.
    
    \item \textbf{Subaru Laser System Update:} Due to the substantial loss in fractional open time compared to Keck 1, Keck 2, and Robo-AO-2, we are in the process of updating the Subaru laser system's keep-out-cone half-angle, which was confirmed to have never been changed from the $2.5^{\circ}$ default.
    
    \item \textbf{Equatorial Observation Challenges:} We identified nearly static closures in sky regions consistent with geosynchronous space assets. Astronomical programs requiring observations near the celestial equator face inherently restricted open times and must absolutely utilize the smallest possible KOC half-angle to be viable.
    
    \item \textbf{Historical Trends and Mega-Constellations:} Analysis of historical FFoV PAMs shows a steady decline in median open time percentages starting around 2016, with a continuing downward trend today. Although the December 2022 waiver for Starlink satellites provided a slight, observable increase in open fractions, the growing volume of low-Earth orbit satellites necessitates additional mitigation strategies.
\end{itemize}

\appendix    

\acknowledgments 

The authors thank the dedicated folks at the Laser Clearinghouse for their help, especially in generating additional PAMs for our experiments. C.B. appreciates the hospitality and support of the Department of Astronomy at Caltech, where much of this work was completed during a sabbatical leave. This study was supported in part by the National Science Foundation under Grant No. AST-2509941.

\bibliography{Baranec_SPIE_2026} 
\bibliographystyle{spiebib} 

\end{document}